\documentclass[authoryear,preprint,review,12pt]{elsarticle}

\usepackage{epsfig}
\usepackage[authoryear]{natbib}
\usepackage{subcaption}
\usepackage{hyperref}
\usepackage{float}

\usepackage{amssymb}

\usepackage{todonotes}
\usepackage{microtype}                 

\usepackage{todonotes}

\journal{IJHCS}

\begin{document}

\begin{frontmatter}



\title{Designing and Evaluating an Adaptive Virtual Reality System using EEG Frequencies to Balance  Internal and External Attention States}


\author[label1]{Francesco Chiossi}

\affiliation[label1]{organization={LMU Munich},
            addressline={Frauenlobstr. 7a}, 
            city={Munich},
            postcode={80337}, 
            state={Bavaria},
            country={Germany}}

\author[label1]{Changkun Ou}
\author[label1]{Carolina Gerhardt}
\author[label2]{Felix Putze}
\author[label1]{Sven Mayer}

\affiliation[label2]{organization={Cognitive Systems Lab, University of Bremen},
            addressline={Enrique-Schmidt-Str. 5}, 
            city={Bremen},
            postcode={28359}, 
            state={Bremen},
            country={Germany}}

\begin{abstract}
Virtual reality finds various applications in productivity, entertainment, and training scenarios requiring working memory and attentional resources.  Working memory relies on prioritizing relevant information and suppressing irrelevant information through internal attention, which is fundamental for successful task performance and training.
Today, virtual reality systems do not account for the impact of working memory loads resulting in over or under-stimulation.
In this work, we designed an adaptive system based on EEG correlates of external and internal attention to support working memory task performance. Here, participants engaged in a visual working memory N-Back task, and we adapted the visual complexity of distracting surrounding elements. 
Our study first demonstrated the feasibility of EEG frontal theta and parietal alpha frequency bands for dynamic visual complexity adjustments. Second, our adaptive system showed improved task performance and diminished perceived workload compared to a reverse adaptation. 
Our results show the effectiveness of the proposed adaptive system, allowing for the optimization of distracting elements in high-demanding conditions. Adaptive systems based on alpha and theta frequency bands allow for the regulation of attentional and executive resources to keep users engaged in a task without resulting in cognitive overload. 

\end{abstract}


\begin{highlights}
\item Developed a VR adaptive system utilizing EEG correlates of external and internal attention to optimizing task performance and user engagement.
\item Demonstrated the effectiveness of online adaptation using EEG correlates of attention, resulting in efficient user model.
\item We adapted peripheral environmental factors rather than manipulating main task features, leading to subtle and natural adaptations that improve task performance.
\item Achieved an accuracy of 86.49\% in classifying internal and external attention states using an LDA model trained on selected EEG frequency bands features.
\end{highlights}    


\end{frontmatter}


\section{Introduction}
The immersive nature of Virtual Reality (VR) environments allows users to engage with a wide range of realistic scenarios, making it an ideal tool for various applications, such as remote collaboration \citep{knierim2021nomadic}, training \citep{zahabi2020adaptive} and entertainment~\citep{lecuyer2008brain}. Here, productivity settings benefited from specific VR applications. However, VR environments' inherent predominant visual nature can challenge users' capacity to process information. For example, users have been overwhelmed when the VR system provided excessive visual stimuli for training in visual tasks \citep{ragan2015effects}, spatial memory \citep{huang2020effects}, and immersive analytics \citep{bacim2013effects, goncalves2022systematic}.

Researchers have proposed adaptive systems that aim to detect if a user is overwhelmed and adjust the VR environment. One promising approach to detect such \textit{overload} states is to employ physiological measures, potentially allowing for online adaptation. A potentially robust approach is detecting the relationship between internally-oriented~\citep{hutchinson2012memory} and externally-oriented~\citep{jiang2021} attention using EEG. 
This is specifically relevant, as many VR tasks can share internal attention~\citep{rowe2000prefrontal, magosso2019eeg} and external attention features~\citep{ricci2022relationship, magosso2019eeg}.  
Users might become overwhelmed, distracted and lose focus if external attention is
prioritized over internal attention. On the other hand, users may miss important external cues if they are in a predominant internal attention state, leading to suboptimal performance in VR. Thus, optimizing internal and external attention processes in VR settings is crucial. Therefore, it is not a matter of whether a task employs internal or external attention exclusively but rather how to optimize one in the face of the other. This is consistent with Chun's taxonomy that internal and external attention are part of a continuum \cite{chun2011taxonomy}.

Internal and external attention share specific EEG features in this continuum~\citep{putze2016, benedek2014alpha, cona2020theta}, i.e., alpha and theta frequency bands. Alpha power is associated with enhancing relevant sensory information processing and concurrent suppression of irrelevant information. Theta mediates Working Memory (WM) and cognitive control processes~\citep{pastotter2013dynamic}. In the context of VR systems, they have also been associated with increased immersion and engagement with the VR task~\citep{magosso2019eeg, ricci2022relationship}. Previous work has employed alpha and theta frequency bands in adaptive VR systems focused on neurofeedback for concentration~\citep{kosunen2016relaworld}, cognitive training~\citep{dey2019} and immersion enhancement \citep{wozniak2021enhancing}. However, most physiologically-adaptive systems focused on main task features, such as visual search targets \citep{dey2019}, learning material \citep{walter2017online} or secondary task difficulty \citep{chiossi2022virtual}. The most closely related to our work is the paper by \citet{vortmann2019eeg}, that, even if in Augmented Reality (AR), used alpha and theta bands for offline classification of internal and external attention states. Thus, the next step is to employ alpha and theta bands for online adaptation of distracting features of the environment from calibrating and allocating the user's attention.

In this work, we designed and evaluated two VR adaptive systems based on EEG correlates of external and internal attention, i.e., alpha and theta frequency bands, that either optimize for external attention or internal attention. We engaged participants in a visual WM task recruiting mostly internal, but also external and attentional resources while adapting the peripheral amount of visual distractors based on the detected attentional state.
We showed how optimizing for internal attention when engaged in a WM task allows for dynamic adaptations of visual distractors, resulting in an increased WM task performance.  We interpreted this as we supported users in avoiding distracting external attention states while remaining engaged with the virtual environment and maintaining an optimal internal attention state.
We make the following contributions: we designed a VR adaptive system that employs visual complexity to dynamically support task performance and engagement (I) and showed that online adaptation of EEG correlates of external and internal attention results in efficient user modeling (II). Third, our study offers a new contribution by focusing on adapting peripheral environmental factors without manipulating main task features using  EEG correlates of internal and external attention (III). Fourth,  based on selected EEG frequency features, we trained a Linear Discriminant Analysis (LDA) model \citep{vortmann2021, wang2007brain} and classified participants' internal and external attention states with an accuracy of $86.49 \%$ (IV). Finally, we make the VR adaptive system openly available with a recorded dataset of behavioral, qualitative, and EEG data (V).

\section{Related Work}
In the following, we highlight the relevance of investigating internal and external attentional states for VR, and then we discuss their EEG correlates in terms of alpha and theta frequency bands. Finally, we summarize previous work that employed EEG as input for adaptation in VR.

\subsection{Relevance of Internal and  External Attentional States in VR}
\label{ssec:RW1}
When immersed in VR, our senses are constantly stimulated, allowing us to interact with the virtual experience. Sensory inputs such as vision and hearing have a strong influence on attention. Still, visual stimulation received the most interest~\citep{hutmacher2019there}, and it is specifically relevant for VR systems, as it is the most predominant stimulated channel in VR~\citep{hvass2017} and can influence the orientation of human attention~\citep{souza2021attention}.  Attention orienting can be described according to the taxonomy proposed by \citet{chun2011taxonomy} for external and internal attention. 

External attention is drawn to external stimuli. Task demands can voluntarily drive external attention in a top-down manner, such as when we focus on a specific spatial location or feature of sensory stimuli that is goal-relevant~\citep{verschooren2019switching}. Alternately, external attention can be captured involuntarily by an object or event in a bottom-up manner, even without specific intention of attending to them~\citep{cona2020theta}. 

Internal attention reflects the processing of internal representations of information. For example, retrieving information about recent or past events (episodic memory)~\citep{hutchinson2012memory}, information to be manipulated in WM~\citep{myers2017prioritizing},  and mental imagery or calculation~\citep{putze2016}. 

Currently, HCI research mostly focuses on internal and external attention for investigating levels of immersion and engagement in VR systems. For example, \citet{magosso2019eeg} explored the conflict between external and internal attention when asked to perform a mental arithmetic task (internal attention) and being immersed in a VR environment (external attention). They reported that a highly-detailed VR environment exerted an external attention capture similar to a reading task, as shown in EEG alpha power. Their result was also confirmed in ~\citet{ricci2022relationship}, which reported how exposure to a VR environment increased their attention to the external environment compared to a relaxation state, a task that recruits internal attention resources. 

Attentional states also influence how much users can be engaged with a specific task. \citet{katahira2018eeg} investigated different flow experiences in an internal attention task, i.e., mental arithmetic task ~\citep{putze2016}, and found that EEG correlates of external and internal attention discriminated between states of overload, boredom and flow. Thus, investigating the external and internal attentional state can benefit users' level of immersion~\citep{souza2021attention} and task engagement~\citep{katahira2018eeg}. 

However, it is important to state that external and internal attention rather than independent states are part of a continuum~\citep{chun2011taxonomy}.  
The continuum between external and internal attention provides a fertile ground for developing adaptive systems. This aspect is specifically relevant for settings where the visual components prevail, such as VR. In particular, the visual nature of VR environments makes it challenging to direct internal attention but also creates opportunities to guide external attention. By leveraging this continuum, adaptive systems can tailor the VR experience to the user's attentional needs and goals, supporting them in achieving optimal performance, immersion and engagement. 

\subsection{Alpha and Theta Frequency Bands as an EEG Correlate of External and Internal Attention}
\label{ssec:RW2}

A large number of studies investigated neurophysiological mechanisms underlying external and internal attention. EEG studies, in particular, have strongly supported the functional significance of two brain oscillatory rhythms: theta (4-8 Hz) and alpha (8-12 Hz). Variations in external and internal attention states are strongly linked to the modulation of alpha and theta frequency bands. These relationships highlight the importance of understanding the underlying mechanisms and their implications for designing efficient  VR adaptive systems that are grounded in physiological inference~\citep{allanson2004research}.

The alpha rhythm is the dominant oscillatory rhythm of the human brain and is traditionally linked to attentional load changes~\citep{foxe2011role}. Alpha band power is thought to act as a sensory gating mechanism by enhancing relevant sensory information processing and suppressing irrelevant information processing~\citep{jensen2010shaping, foxe2011role}. Thus, alpha activity plays a crucial role in regulating attention processes, both within and outside the focus of attention. Studies have explored posterior alpha as a possible index of internal and external attention, with external attention linked to alpha power decrease and internally directed attention primarily associated with alpha power increase~\citep{cona2020theta, benedek2014alpha}.
Specifically, alpha increase aims at preventing external, irrelevant sensory information from interfering with internal processes. On the other hand, when individuals enter an external attention state, alpha power tends to decrease in the occipital region. This decrease in alpha frequency band reflects increased excitability of the visual cortex, which in turn enhances the processing of external sensory information~\citep{van2019functional}.

Regarding theta frequency band, its increased power has been linked to WM engagement and cognitive control, particularly in frontal regions~\citep{harmony2013functional}. Both WM and cognitive control involve internal attention features. WM requires temporarily maintaining and manipulating internal representations of information~\citep{rerko2013focused}. Cognitive control refers to the ability to regulate thoughts and actions to achieve specific goals~\citep{braver2012variable}, and therefore ignoring task-irrelevant or distracting information~\citep{lavie2010attention}. The theta activity could be indicative of a balance between external and internal attention~\citep{cona2020theta} and their competition~\citep{magosso2021alpha}. Theta decrease may signify the act of shifting attention towards external stimuli, allowing for the processing of potentially distracting information. In contrast, an increase in frontal theta underlies protection and prioritization of ongoing internal processing ~\citep{lorenc2021distraction, de2020oscillatory}. 

In conclusion, alpha and theta changes can index different levels of the continuum between external and internal attention, namely, their competition. In the next section, we review adaptive and passive BCI (pBCI) systems that employ such frequency bands as input for VR systems.


\subsection{EEG as an Input for Adaptation in Virtual Reality}
\label{ssec:RW3}
EEG frequency bands have been used as the primary input for interaction in pBCI systems. A pBCI system derives an output from automatic, involuntary, spontaneous brain activity, interpreted in the given context~\citep{lotte2018review}. Historically designed for communication and control for patients with severe disabilities,  pBCIs recently found new applications for patients and healthy users when combined in VR settings~\citep{lecuyer2008brain}. pBCIs and VR can see reciprocal benefits as pBCI can become more intuitive than traditional devices. At the same time, VR can enrich interaction and provide more motivating feedback for pBCI users than traditional desktop settings~\citep{arico2018passive}. Therefore, VR-pBCI or physiologically-adaptive VR systems could support system learnability, i.e.,  reduced time required to learn BCI skills or increased classification performance~\citep{leeb2006walking, ron2009brain}, and allow for an extensive range of applications~\citep{chiossi2022it}.


In VR adaptive systems, alpha and theta were the basis for designing adaptive systems for meditation~\citep{kosunen2017neuro} and adaptation of task difficulty based on cognitive interference~\citep{wu2010optimal}. Another related work focused specifically on alpha for cognitive training is the study by \citet{dey2019}, where authors modulated the visual task difficulty in a VR visual search task. Finally, frontal theta power has also been employed in adaptive systems to index the continuum between overload and optimal motivational engagement~\citep{ewing2016evaluation}. Closer to our work, even though applied to AR settings, are the adaptive systems developed by \cite{vortmann2022atawar, vortmann2020attention}. Here, the authors employed the entire EEG frequency spectrum and eye tracking to categorize internal and external attention with an 85.37 \% accuracy in a special alignment task.

Previous work explored alpha and theta EEG frequencies for adaptation in human factors, VR and AR environments, but mostly for interaction methods and monitoring cognitive load or task engagement. However, only a few works investigated the use of EEG for external and internal attention in VR settings~\citep{magosso2019eeg,magosso2021alpha} and adaptive systems have been designed only in AR settings~\citep{vortmann2019eeg}. Our research is the first that investigates how to develop an adaptive VR system to optimize for internal attention, grounded in physiological inference~\citep{allanson2004research}, and validated in a user study.

\subsection{Summary}
The immersive nature of VR technology has revolutionized how we interact with digital content. However, VR is primarily designed around visual information that challenges users' capacity to process information ~\citep{bacim2013effects, goncalves2022systematic}, leading to an unbalanced allocation of external attention resources at the expense of internal attention ~\citep{vortmann2021}. Thus,  the design of an adaptive VR system grounded in EEG correlates of external/internal attention state, leveraging the amount of task-irrelevant elements in the internal-external attention continuum ~\citep{chun2011taxonomy}, can impact subjective workload, engagement, and task performance ~\citep{arico2018passive}. Here, we compare two adaptive systems, one optimizing for external attention (\textsc{Negative Adaptation}) and one for internal attention (\textsc{Positive Adaptation}) while participants engaged in a visual N-Back, a task that primarily recruits internal attention but also features both attention components, which act along a spectrum.
Based on related work, we designed an adaptive system to support performance by balancing the two attentional components. However, given the inherent trade-off, we designed two systems that optimize for external or internal attention, we hypothesize that: 

\begin{enumerate}
     \item [\textbf{HP1}:] An adaptive system designed for balancing the attention competition towards internal attention should positively impact WM task performance.
     
     \item [\textbf{HP2}:] An adaptive system designed for balancing the attention competition towards external attention should negatively impact WM task performance.
     
     \item [\textbf{HP3}:] By optimizing the visual complexity and achieving a balanced allocation of internal and external attention resources, the adaptive system designed for internal attention is hypothesized to increase subjective engagement in the WM task.
     
     \item [\textbf{HP4}:] If the adaptive system balancing for external attention has a detrimental effect on WM task performance, we expect increased subjective workload ratings.
 \end{enumerate}

Moreover, detecting and understanding a user's attentional state could significantly enhance the utility of VR systems and enable novel use cases that are purposefully designed to react, detect and optimize it~\citep{allanson2004research}. Therefore,  drawing from AR settings \citep{vortmann2019eeg, vortmann2021} and considering how much internal and external attention are recruited in VR settings, we expect that:

\begin{enumerate}
     \item [\textbf{HP5}:] External and internal attentional states in VR can be reliably classified using an LDA model \citep{wang2007brain, vidaurre2010toward}.
 \end{enumerate}

We explore classification-based differentiation of external and internal states as an alternative to literature-driven selection of adaptation variables from the EEG signal. Potentially, machine learning can better balance multiple such variables in one model and deal better with EEG trial-by-trial fluctuations \citep{lotte2018review}. As this approach requires more tuning and is less predictable, we explore its potential for future adaptation approaches.
 
\section{Architecture of the EEG-Adaptive VR System}
\label{ssec:arch_ada}

VR environments are often designed to be immersive, realistic, and engaging, making it easy for users to become distracted or overwhelmed by external visual stimuli. Thus, we might see a constant competition between internal and external attention when engaged in VR scenarios. Here, an EEG-adaptive system can monitor users' attentional states and optimize attentional processing to improve internal task performance in VR settings by adapting surrounding visual information. We define the goal of optimizing attentional processing as enhancing the efficiency and effectiveness of attentional processing necessary for a given task. This goal requires identifying and achieving an ideal balance between external and internal attentional processes to improve task performance while maintaining engagement with the virtual environment. The critical aspect is not whether a task exclusively relies on internal or external attention, but rather how to achieve an optimal balance between the two. For example, during a mostly internal task, the goal is to provide external attention as much as possible without compromising the focus on the internal processing of the task. This aligns with \citet{chun2011taxonomy} perspective that internal and external attention are interconnected along a continuum, and their interaction must be considered when optimizing attentional processing.

In this work, we designed and compared two VR adaptive systems based on EEG correlates of internal and external attention. We frame the adaptive systems from the perspective of a situation in which being in a state of internal attention is desirable. Specifically, the system called from here on \textsc{Positive Adaptation} is designed to optimize the internal attention state. In contrast, the system defined as \textsc{Negative Adaptation} aims to optimize externally-directed attention. We used the visual WM N-Back task developed by \citet{chiossi2022virtual} for both adaptive systems.  We chose the VR N-Back task as it recruits WM resources and results in changes in alpha and theta frequency bands~\citep{chiossi2023exploring, tremmel2019estimating}. We adapted the surrounding visual complexity of the VR environment in the form of non-player characters (NPCs) that were passing next to the participant. We denote the number of NPCs passing by the participants per minute as  \textsc{Stream}. The \textsc{Stream} of NPCs was constant, making NPCs appearing/disappearing at the same rate. The \textsc{Stream} of NPCs contributes to the general amount of detail, clutter and objects in the scene, namely its visual complexity~\citep{olivia2004identifying}. NPCs are task-irrelevant elements, and for the purpose of this task, they act as distractors.

\subsection{EEG Adaptive System}
\label{Adaptation_system}
Both adaptive systems shared the same apparatus encompassing four components:
(I) an R-Net 64 channel EEG with two wireless LiveAmp amplifiers (BrainProducts, Germany), (II) Transmission Control
Protocol (TCP)/Internet Protocol (IP) for online EEG data preprocessing (III) the Unity 3D (Version  2022.1) game engine for VR development; and (IV) HTC Vive Pro (HTC, Taiwan) VR HMD for the display of the VR environment. For online adaptation, we first applied a notch filter at 50 Hz and then performed a band-pass filtering between (1-70 Hz) to remove high and low-frequency noise. Then, we extracted alpha and theta EEG powers via Welch's periodogram method using a Hamming window
of 5 seconds at 50\% overlap, zero-padded to 10 s, to obtain a 0.1 Hz frequency resolution. For determining the alpha frequency range, we computed the Individual Alpha Frequency (IAF) via the method developed by Corcoran et al.~\citep{corcoran2018toward}. This allowed us to identify an individualized alpha range for each participant. Then, based on the individual alpha lower bound, we defined the theta frequency range, using the alpha lower bound as the high theta bound and defining the theta lower bound by subtracting 4 Hz from the alpha lower bound. For computing alpha power  we used  parieto-occipital channels (P3, Pz, PO3, POz , PO4, O1, O2)~\citep{benedek2014alpha, magosso2019eeg}, while for theta, we chose frontal channels (Fp1 , Fp2, AF3, AF4, F1, F2, F3, Fz, F4, FC1, FC2)~\citep{magosso2019eeg}. Electrode FCz was set as an online reference.

For data streaming and online preprocessing, we transmitted the data through a Transmission Control Protocol (TCP)/Internet Protocol (IP) client to a TCP/IP server implemented via Python network programming. This implementation enabled us to exchange data between Lab Streaming Layer\footnote{\url{https://labstreaminglayer.org/}} and the VR Unity environment in both forward and backward directions. We utilized a Network Time Protocol (NTP) service to time-synchronize the VR Unity scene's time and the bridge server's operating system time.

\begin{figure}[t]
    \subfloat[Positive Adaptation]
    {%
        \includegraphics[width=.485\textwidth]{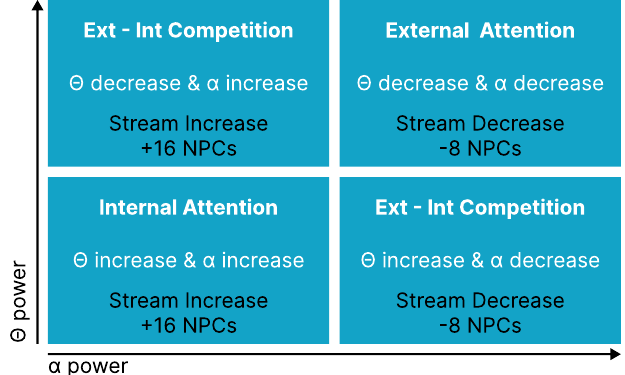}
    \label{fig:good_matrix}
    }
    \subfloat[Negative Adaptation]
    {%
        \includegraphics[width=.485\textwidth]{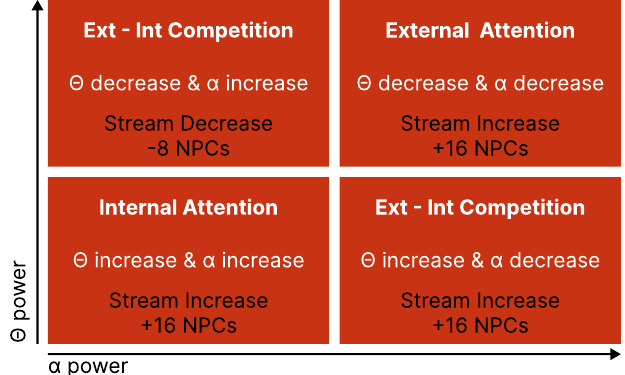}
        \label{fig:bad_matrix}
    }
    \caption{Adaptation Methodology for the two adaptive systems based on the increase and decrease of the alpha and theta frequency bands and their relevance to internal and external attentional states.}
    \label{fig:matrix}
\end{figure}

\begin{figure}
    \centering
    \includegraphics[width=\textwidth]{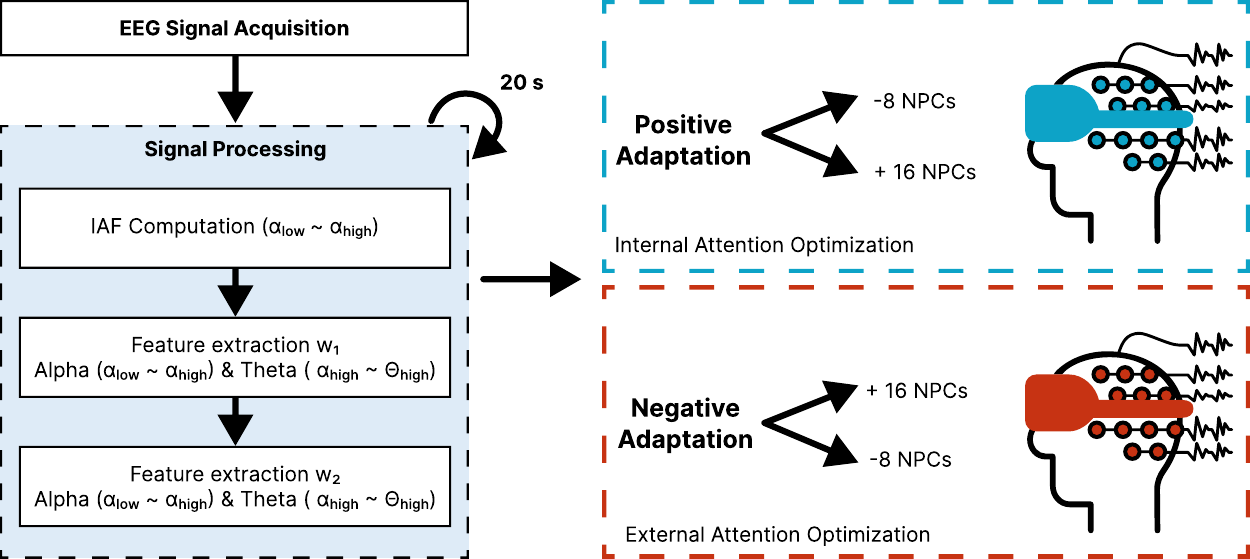}
    \caption{Architecture of the two adaptive systems. The Stream of NPCs adapts based on alpha and theta variation in two different time windows ($w_1$ and $w_2$), each lasting 20s. If the change is bigger than the decision threshold of  15\%, the NPC stream is either increased by +16 or decreased by -8 NPCs. The Positive Adaptation system (a) aims at optimizing internal attention, while the Negative Adaptation system (b) targets external attention.}
    \label{fig:architecture}
\end{figure}

%
%

\begin{figure}[htbp]  
    \centering  
    \subfloat[Visual Monitoring]{%
        \includegraphics[width=.46\linewidth]{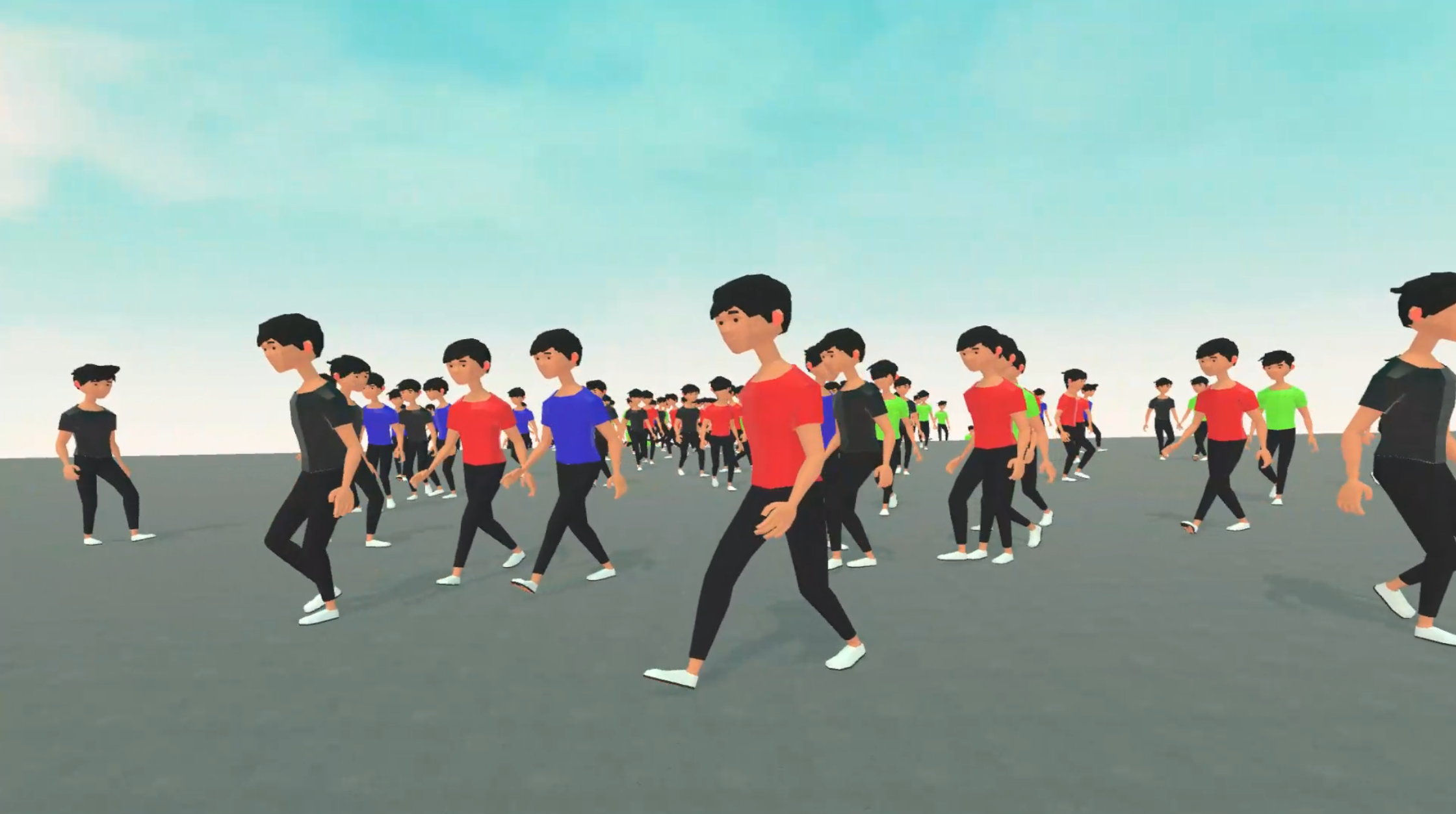}
        \label{fig:task_VisMon}
    }
    \hfill  
    \subfloat[N-Back No Adaptation]{%
        \includegraphics[width=.46\linewidth]{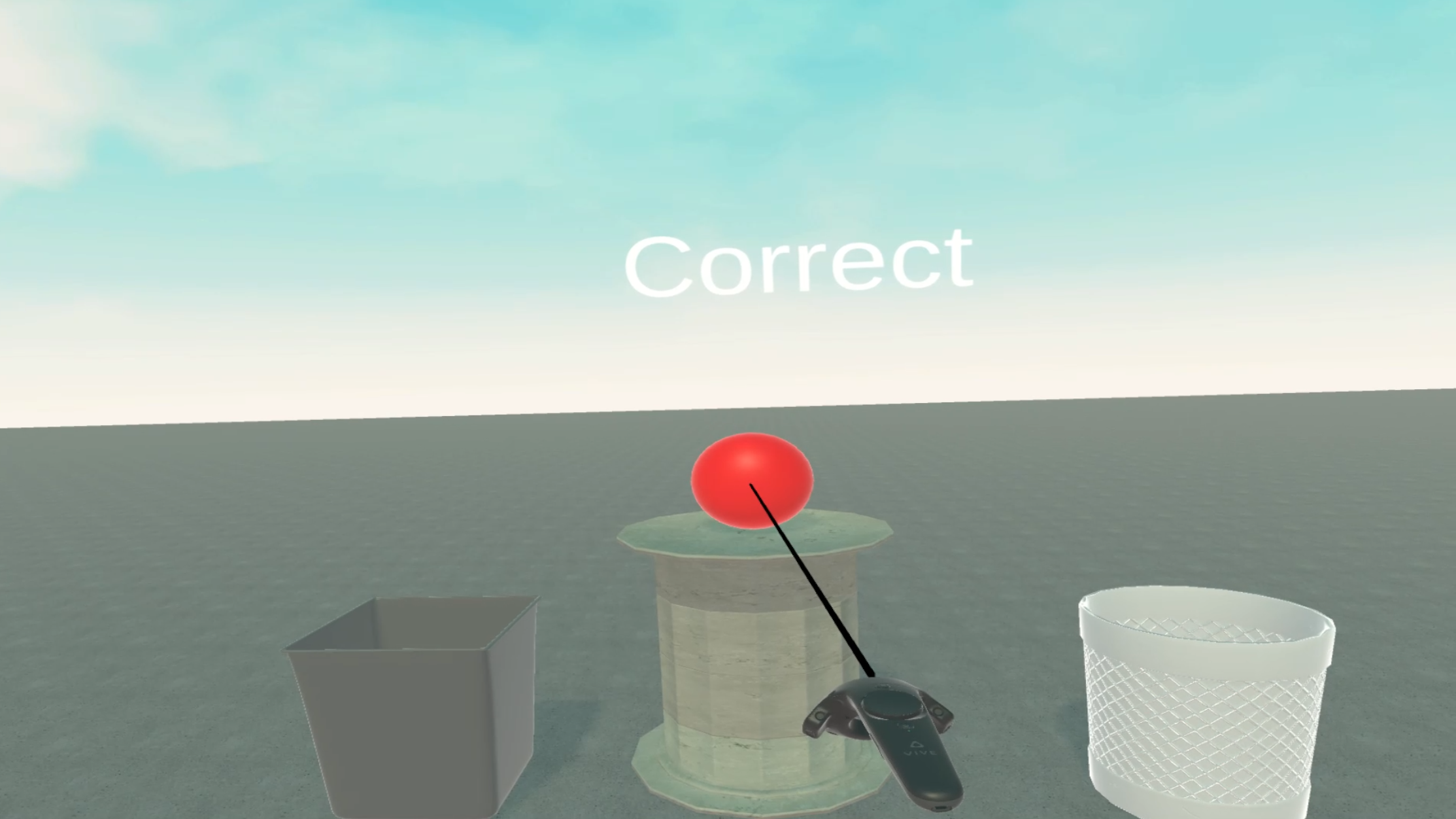}
        \label{fig:task_Nback_noadapt}
    }
    
    \subfloat[N-Back Positive Adaptation]{%
        \includegraphics[width=.46\linewidth]{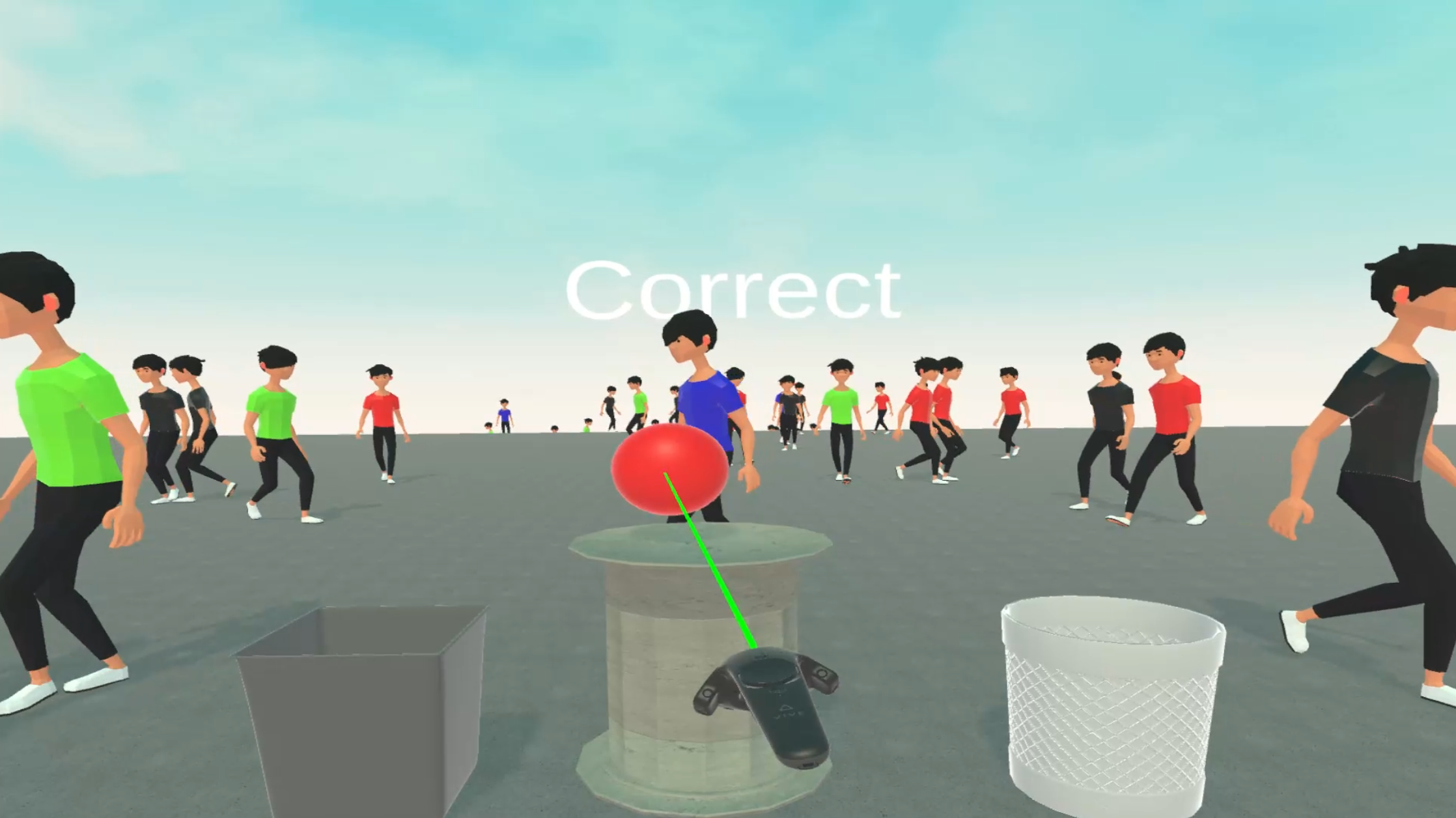}
        \label{fig:task_Nback_pos_adapt}
    }
    \hfill
    \subfloat[N-Back Negative Adaptation]{%
        \includegraphics[width=.46\linewidth]{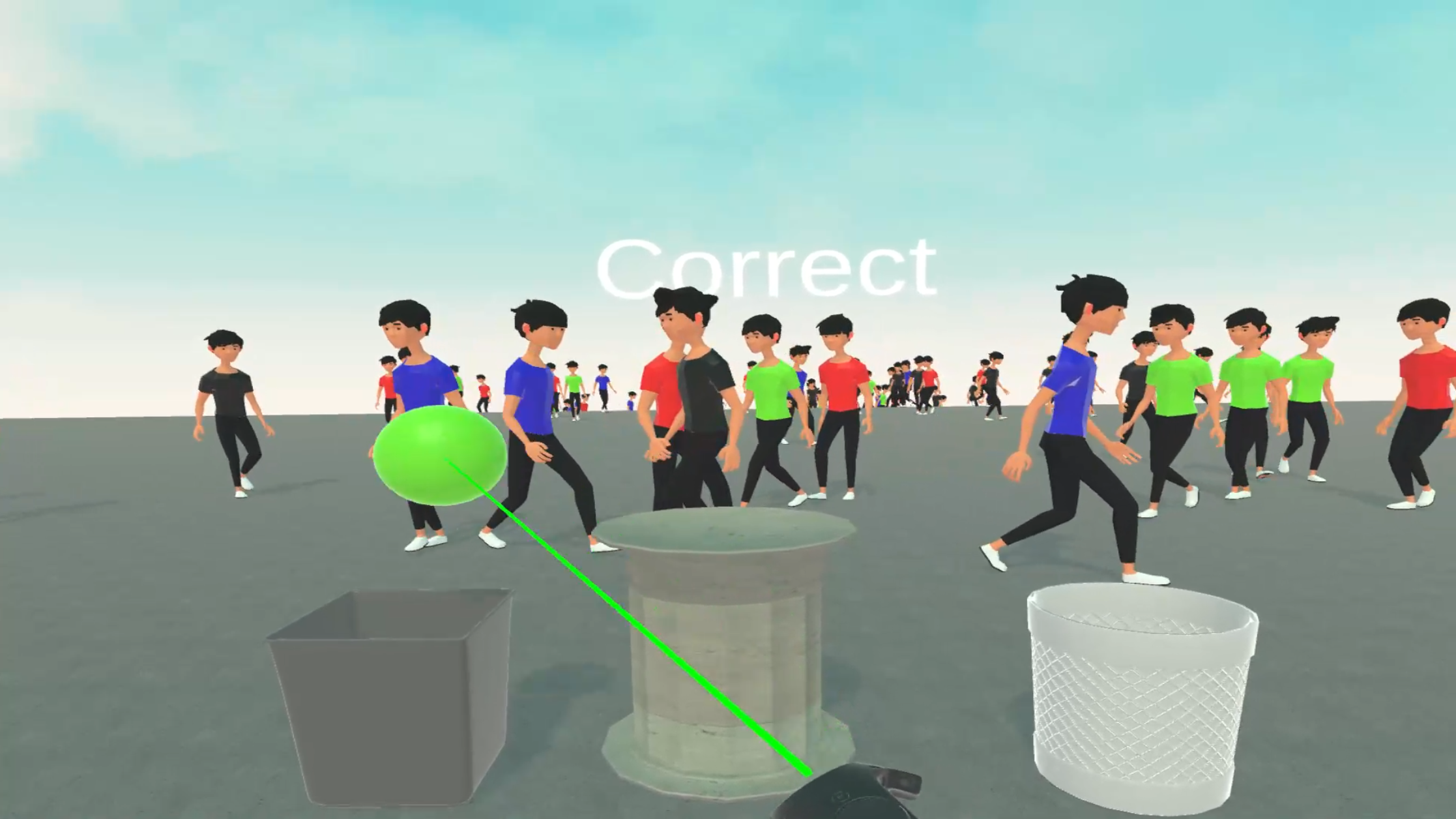}
        \label{fig:task_Nback_neg_adapt}
    }
    \caption{Game VR Capture of the experimental tasks. In the Visual Monitoring task (a), participants were exposed to a \textsc{Stream} of NPCs and asked to monitor, i.e., follow with their gaze NPCs with a specific colour. In the N-Back No Adaptation (b), participants actively interact with a sequence of spheres presented on a marble-like pillar and have to place them into either the left or right bucket. The placement of each sphere is determined by its color, and the sphere's color presented two steps prior (N=2). The sphere has to be placed on the left if the color is different and on the right bucket, if the color is the same.}
    \label{fig:Stream_Visualization}
\end{figure}

\subsection{Adaptive System Architecture}
Adaptive system architecture was grounded on previous work on the functional significance of alpha and theta frequency bands~\citep{putze2016, benedek2014alpha, vortmann2019eeg} as input for the VR adaptive systems. First, we used a continuous adaptation, continuously comparing the mean alpha and theta bands over two consecutive time windows, $w_1$ and $w_2$, both of 20 seconds duration , based on previous work \cite{chiossi2022virtual,chiossi2023exploring}. Second, we compute the mean alpha and theta power for $w_1$ and $w_2$. Here, we compare the direction of change (defined as exceeding a 15 \% threshold) of both mean alpha and theta in $w_2$ to the average power in $w_1$. We determined the threshold after multiple sessions (N=14, $M=25.62$, $SD=2.52$; 7 female, 7 male, none diverse) to identify a threshold allowing the system to optimize external attention while avoiding overshooting, i.e., always performing the same adaptation response or undershooting, i.e., not reacting to changes in alpha and theta EEG frequencies. We tested multiple thresholds (5\% steps from 5-30\%) and evaluated system performance. If the change from $w_1$ to $w_2$ of both alpha and theta exceeded the decision threshold, depending on the direction of the frequency band, a change in \textsc{Stream} of NPC is performed. We define our optimization goal as  biased towards that specific type of attention, but still tries to maintain a certain balance.

\begin{figure}
\centering
    \includegraphics[width=.8\textwidth]{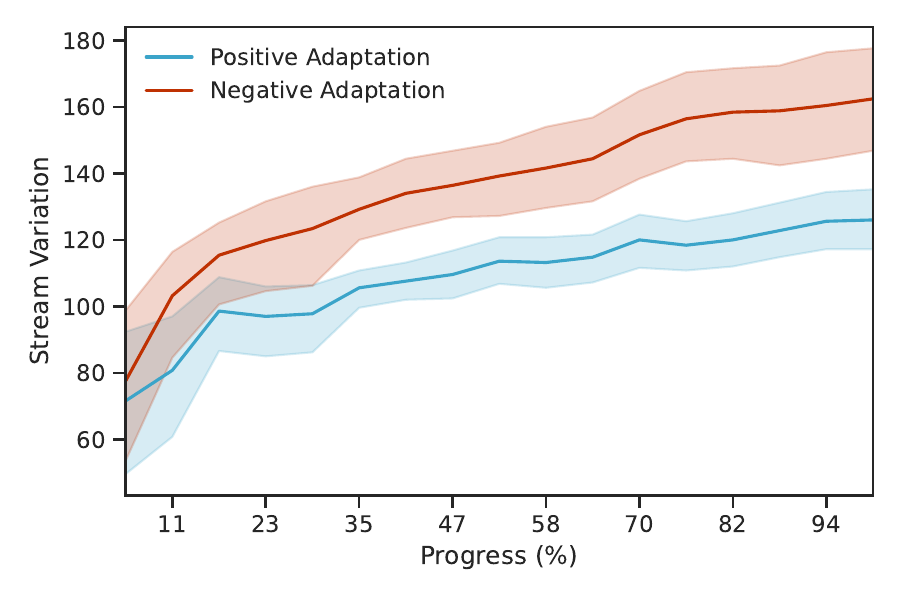}
    \caption{Stream Visualization. Here, we depict the average evolution over time of the \textsc{Stream} for the two adaptive systems. The \textsc{Positive Adaptation} averaged on $133.17$ NPCs per minute while the \textsc{Negative Adaptation} on $161.48$ NPCs.}
    \label{fig:stream}
\end{figure}

In the \textsc{Positive Adaptation} system, when a shared 15\% increase in both alpha and theta is detected in $w_2$, as compared to the previous 20 s in $w_1$, the user is assumed to be in an internal attention state, therefore to find an optimal level of visual complexity, the system increases the \textsc{Stream} by 16 NPCs . By doing so, we aim to test the tradeoff between internal attention and external visual complexity. This approach allows us to investigate how individuals adapt to a dynamic environment where attentional demands are subject to change. In the opposite case, when alpha and theta decrease by at least 15 \% , participants are assumed to be in an external attention state. Therefore, 8 NPCs are removed from the scene to support the internal attention state. This decision tree is grounded on the fact that internal attention is associated with an increase in alpha~\citep{benedek2014alpha,connell2009uncovering} and theta~\citep{cona2020theta}, and reflecting increased WM engagement~\citep{de2020oscillatory}. Alternately, theta and alpha can show opposite directions. When alpha decreases and theta increases by 15\%, we assume that users entered an external attention state as indexed from alpha band~\citep{benedek2014alpha}, and increased cognitive control~\citep{braver2012variable} due to the increased effort to maintain the focus while ignoring the distractors. In this case, the \textsc{Stream} is decreased by 8 distractors. In contrast, if alpha increases and theta decreases, we theorize an increase in internal attention and a decrease in WM engagement. Therefore, we increase the \textsc{Stream} by adding 16 distractors. Those parameters are based on previous work on adaptive system design accounting for the task irrelevance and distracting effect of the NPCs \cite{chiossi2023exploring, chiossi2023adapting}.  Secondly, they allow to avoid the numbers of distracting NPCs drops to 0 per minute. Participants started the adaptive blocks with a \textsc{Stream} set at 115 NPCs entering the scene per minute. On average, \textsc{Stream} in \textsc{Positive Adaptation} condition stabilized  on $133.17$ NPCs per minute ($SD=14.86$), see\autoref{fig:stream}. Participants executed a mean of 152.25 ($SD=73.19$) WM trials in \textsc{Positive Adaptation}, 167.33 ($SD=68.04$), in \textsc{Negative Adaptation} 182.96 ($SD=68.53$). The \textsc{Positive Adaptation} methodology is depicted in \autoref{fig:good_matrix} and the architecture in \autoref{fig:architecture}.

The \textsc{Negative Adaptation} system, optimizing for external attention, follows a different strategy. Thus, the \textsc{Stream} is increased when the participants are detected in a state of external attention or when there is a decrease of at least 15\% in alpha power in $w_2$ compared to $w_1$ (Ext-Int competition), pointing towards an increased state of external attention.
When alpha and theta bands show an increase of 15\% in $w_2$, and therefore in a state of internal attention, the \textsc{Stream} is increased to drive participants in a higher visual complexity environment and increase their external attention state. When the participants are in a state of external-internal competition, i.e., alpha power increases and theta decreases, reflecting an increase in internal attention but a decreased WM engagement, the \textsc{Stream} is decreased by - t8 NPCs to drive them in a state of boredom, as previously designed by \citet{ewing2016evaluation}. This choice is meant to evaluate if adaptation can still impact the user's WM performance without improving it, demonstrating that BCI-based adaptation cannot be replaced equivalently with a purely performance-based one. If participants already exhibit an internal focus of attention, this might decrease engagement with the task, enforcing such an internal state. Finally, when alpha and theta have the same direction, indexing an internal attention state, the system increases the visual complexity by adding 16 NPCs to the \textsc{Stream}.  On average, \textsc{Stream} in \textsc{Negative Adaptation} condition stabilized on $161.48$ NPCs per minute ($SD=21.8$). The \textsc{Negative Adaptation} methodology is depicted in \autoref{fig:bad_matrix} and the architecture in \autoref{fig:architecture}.

\begin{figure}[t]
    \centering
    \includegraphics[width=\linewidth]{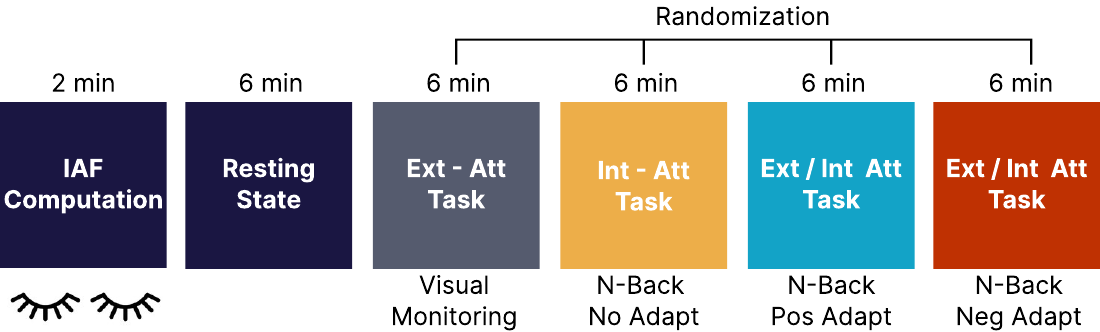}
    \caption{\emph{Experiment Procedure}. The experiment encompassed six different blocks. In between blocks, participants filled in NASA-TLX and GEQ subscales and observed a three-minute pause in VR. Blocks order was randomized for the Visual Monitoring, N-Back with No Adaptation and N-Back with Positive or Negative Adaptation. In the first block, participants maintained their eyes closed to compute the Individual Alpha Frequency (IAF). In the Resting state block, participants relaxed in the neutral VR environment with distracting elements. After those two blocks, participants experienced the experimental tasks (Visual Monitoring, N-Back No Adaptation,  N-Back Positive Adaptation,  N-Back Negative Adaptation block) in a randomized order. Refer to \autoref{ssec:arch_ada} for a complete description of the adaptive systems.}
    \label{fig:Task_procedure}
\end{figure}

\section{User Study}

The study evaluated if adaptation of visual complexity, based on EEG correlates of internal and external attention, can optimize behavioral WM performance and subjective engagement ratings compared to a system designed to optimize for external attention. As the main task, we chose the established N-Back task~\citep{soveri2017working} in the VR version as adapted from~\citet{chiossi2022virtual}. The task involved updating the information in working memory and paying continuous attention to the presented spheres while retaining the previously presented information. We selected this task because it evokes external and internal attention processing, making it ideal for optimizing one of the two processes in adaptive systems.

\subsection{Design}
To examine differences in behavioral performance, perceived workload and engagement and alpha and theta frequency bands, we performed a within-subjects study for the system's adaptability factor (\textsc{Positive} vs \textsc{Negative Adaptation}). 
The experiment encompasses six blocks, of which four are the experimental ones and either recruit only external (Ext-Att Task : Visual Monitoring Task) or internal attention (Int-Att task : N-Back No Adaptation), and two adaptive blocks which have a competition between the two processing with two different adaptive systems (Ext / Int Task: N-Back Negative Adaptation and Ext / Int Task: N-Back Positive Adaptation).  The first two blocks are the Individual Alpha Frequency Block (IAF computation Block), which lasted 2 minutes and is necessary for computing the IAF for each participant, and the Resting State block, used as a basal condition for normalization to the experimental blocks.  The Ext-Att Task (Visual Monitoring task) requires participants to inspect the VR scene, identify and follow with the gaze NPCs of a specific color, see \autoref{fig:task_VisMon}.  The Int-Att Task (N-Back No Adaptation) is a visual N-Back task (N=2) where the participants have to retain information regarding the color of a sphere and internally direct attention towards the memory of the color of the sphere and compare it to the color of the current sphere, and place in a specific bucket depending on the match of the color, see \autoref{fig:task_Nback_noadapt}.
The two "adaptive" experimental conditions required participants to perform the N-Back task while being exposed to a \textsc{Stream} of NPCs, i.e, an adaptation of the visual complexity through changes in the participant's alpha and theta EEG frequency bands. In the two adaptive tasks, NPCs serve as distractors as they are elements that are not relevant to the task at hand (see \autoref{fig:task_Nback_pos_adapt} for the Positive Adaptation and see \autoref{fig:task_Nback_neg_adapt} for the Negative Adaptation).
Respectively, positive adjustments of \textsc{Stream} (Increase) resulted in adding 16 NPCs to the scene, while negative adjustments of \textsc{Stream} (Decrease) resulted in removing 8 NPCs from the scene.

\subsection{Task}
Participants executed two types of tasks, i.e., Visual Monitoring task and N-Back task. In the Ext-Att block, participants were exposed to a fixed \textsc{Stream} (334 NPCs per minute) and were asked to monitor and follow with the gaze approaching NPCs of a randomized color (blue, green, black and red). This Visual Monitoring task is expected to recruit external attention resources as it only requires visual processing and externally-directed attention to participants. This block acts as a control condition as it is the only one in which participants performed a task which mainly required external attention.

In the Int-Att Block and in the two adaptive blocks, participants executed the N-Back (N=2) as adapted from \citet{chiossi2022virtual}. Here, participants are presented with a sequence of spheres over a marble-like pillar that has to be placed in one of two buckets on the left and the right, respectively. Spheres could have been spawned in four possible colors (green, red, blue, and black), according to~\cite{mcmillan2007self}, in a randomized sequence. Participants were required to pick up the spheres with an HTC Vive Pro controller and place them in the correct buckets. The placement of each sphere depended on its color and the color of the sphere presented two steps before. If the colors matched, the participant had to place the sphere in the right bucket. If the colors did not match, the participant had to put the sphere in the left bucket. New spheres would appear either after the current sphere was placed in one of the two buckets or after 4 seconds. Participants received accuracy feedback every 20 sphere placements and were instructed to maintain a performance level of 90\%. Errors were computed by the proportion of times the sphere was positioned in the wrong bucket.

\subsection{Procedure}
Upon participants' arrival, we provided them with information regarding the study's procedure and addressed any inquiries they had before having them sign the informed consent form. The study began with a trial phase to enable participants to acclimate to the VR environment. During the VR trial phase, participants practised the 2-back task until they achieved a minimum accuracy level of 95\% while identifying a sequence of 80 spheres \citep{chiossi2022virtual}.
Next, the experimenter set up the water-based EEG cap. The experimental procedure started with the IAF Block, where participants kept their eyes closed for 2 minutes and 10 seconds. We describe the IAF computation in \autoref{sec:IAF_comp} . Then participants observed 3 minutes of rest for physiological adaptation (not included in the analysis) and started the Resting State Block for 6 minutes. They sat comfortably in the VR environment without NPCs or N-Back task elements, keeping their hands on their thighs without moving. 
After the Resting State, participants moved to the experimental phase consisted of four randomized experimental blocks (Ext-Int task, Int-Att Task, Positive Adaptation and Negative Adaptation), lasting six minutes each. In between blocks, participants fill the NASA TLX questionnaire to evaluate perceived workload~\citep{hart1988development} and the Game-Experience Questionnaire (GEQ) In-Core Module, choosing the Competence, Immersion, and Positive Affection subscales for validated content validity for perceived engagement~\citep{law2018}. Immersion and Competence subscales measure the level of engagement participants experience with the task at hand which is related to challenge immersion~\citep{burns2015use}.  Again, between questionnaire completion, participants rest for 3 minutes in the VR scenario for physiological adaptation. Overall, the experiment lasted one hour and thirty minutes.

\subsection{Participants}
We recruited 24 participants ($M=28.5$, $SD=6.06$; 12 female, 12 male, none diverse)  via convenience sampling and social media. However, we removed 2 participants due to technical interferences, resulting in a total population of 22. Participants provided written informed consent before their participation.  None of the participants reported a history of neurological, psychological, or psychiatric symptoms.

\subsection{Offline EEG Recording and Preprocessing}
\label{sec:EEG_pipeline}
EEG data were recorded from 64 Ag-AgCl pin-type passive electrodes mounted over a water-based EEG cap (R-Net, BrainProducts GmbH, Germany) at the following electrode locations: (Fp1, Fz, F3, F7, F9, FC5, FC1, C3, T7, CP5, CP1, Pz, P3, P7, P9, O1, Oz, O2, P10, P8, P4, CP2, CP6, T8, C4, Cz, FC2, FC6, F10, F8, F4, Fp2, AF7, AF3, AFz, F1, F5, FT7, FC3, C1, C5, TP7, CP3, P1, P5, PO7, PO3,
Iz, POz, PO4, PO8, P6, P2, CPz, CP4, TP8, C6, C2, FC4, FT8,
F6, F2, AF4, AF8 according to the 10–20 system. Two LiveAmp amplifiers acquired EEG signals with a sampling rate of 500 Hz. All electrode impedances were kept below $\le$ 20 k$\Omega$. We used FCz as an online reference and AFz as ground. 
For offline preprocessing we used  MNE Python~\citep{gramfort2013meg}. We first notch-filtered at 50 Hz followed by a band-pass filter between 1-70 Hz to eliminate noise at high and low frequencies. Next, we re-referenced the signal to the common average reference (CAR) and applied the Infomax algorithm  for Independent Component Analysis (ICA). We utilized the "ICLabel" MNE plugin~\citep{pion2019iclabel} for automatic classification and correction of ICA components. On average, we removed $2.97$ ($SD=5.19$) independent components within each participant.

\subsection{Individual Alpha  and Theta Frequencies Bands Range Computation}
\label{sec:IAF_comp}
We employed the methodology established by \citet{corcoran2018toward} to calculate IAF, based on \citet{klimesch2012alpha}. This method enables us to determine the alpha band at the individual level, taking into account the differences between individuals, thereby facilitating a more accurate and detailed online adaptation and offline analysis. We removed the first and last four seconds of data from the beginning and end of each IAF recording to remove signals unrelated to cortical activity and impacted by eye blinks. For IAF computation, we use posterior electrodes (P3, Pz, PO3, POz , PO4, O1, O2).  Overall, the lower alpha range stabilized across participants on an average of  8.02  Hz ($SD=.09$), while with the higher bound, we obtained an average of 12.99 Hz ($SD=1.03$). After determining the IAF for each participant, we utilized this information to calculate the alpha power for parieto-occipital electrodes employed for adaptation, see \autoref{Adaptation_system}. For Theta power, we applied to a window of 4 Hz falling below the alpha lower bound computed from the IAF. Participants showed an average theta range of 4.02 Hz ($SD=.09$) - 8.02 ($SD=.09$). We then computed the Theta power from the frontal electrodes selected for adaptation, see \autoref{Adaptation_system}.

\section{Results}
In this section, we first present results on EEG power bands, behavioural accuracy and subjective scores on perceived workload (NASA-TLX) and engagement (GEQ) using Repeated measures ANOVA or Friedman's test for not normally distributed data as evaluated by the Shapiro-Wilk test. For post hoc comparisons, we use Conover's tests with Bonferroni correction. We compared the effect of \textsc{Block} (N-Back No Adaptation, N-Back Positive Adaptation, N-Back Negative Adaptation) over measured dependent variables. For subjective measures, we also include the Visual Monitoring task for comparison. We employ a Generalized Linear Mixed Model (GLMM) for reaction times to investigate differences in the reaction time distributions. Finally, we report our results on the classification of the two attentional states based on Visual Monitoring (External Attention) and N-Back task with No Adaptation (Internal Attention).

\subsection{EEG Results}

\begin{figure}
    \centering
    \includegraphics[width=.7\linewidth]{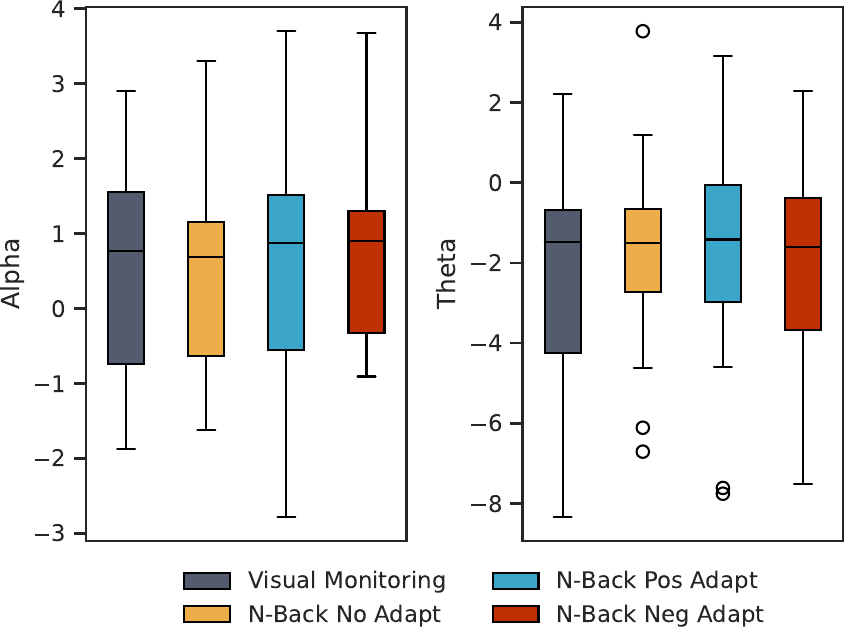}
    \caption{EEG Results. Boxplot representing average Alpha (left) and Theta (right) frequencies. Frequencies were obtained from the parieto-occipital channels for Alpha, while for Theta, we chose frontal channels. Values are computed for each experimental condition and normalized to the resting state.}
    \label{fig:results_eeg}
\end{figure}

\begin{figure}[h]
    \subfloat[]{%
        \includegraphics[width=.23\textwidth]{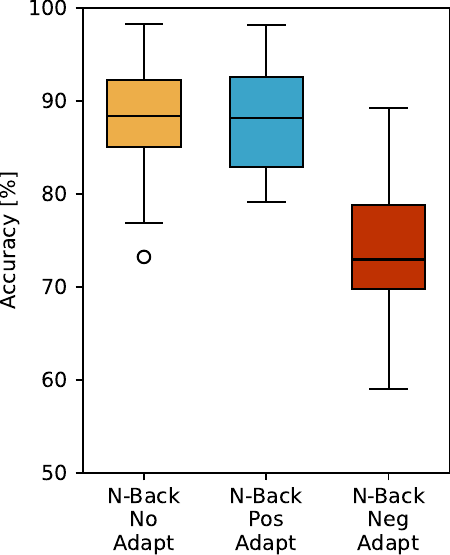}
    \label{fig:results_acc}
    }
    \hfill
    \subfloat[]{%
        \includegraphics[width=.72\textwidth]{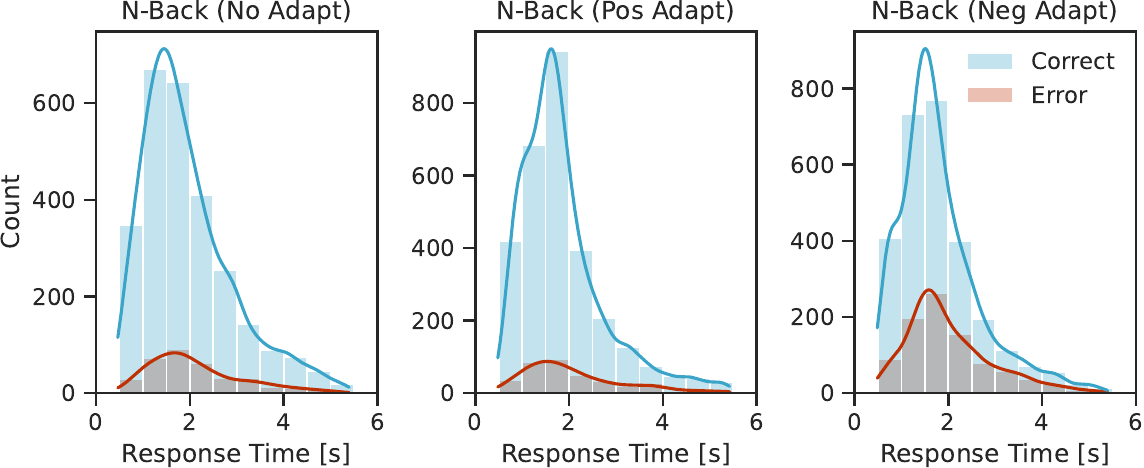}
        \label{fig:results_rt}
    }
    \caption{Behavioral Results. On the left (a), we present the results on Behavioral Accuracy. Here, participants significantly showed higher accuracy in N-Back and Positive Adaptation conditions as compared to the Negative Adaptation. On the right (b), we present an overview of reaction time distributions, separated by correct and error responses. No significant differences were detected in reaction times distributions.}
\end{figure}

\begin{figure}[t]
    \centering
    \includegraphics[width=\textwidth]{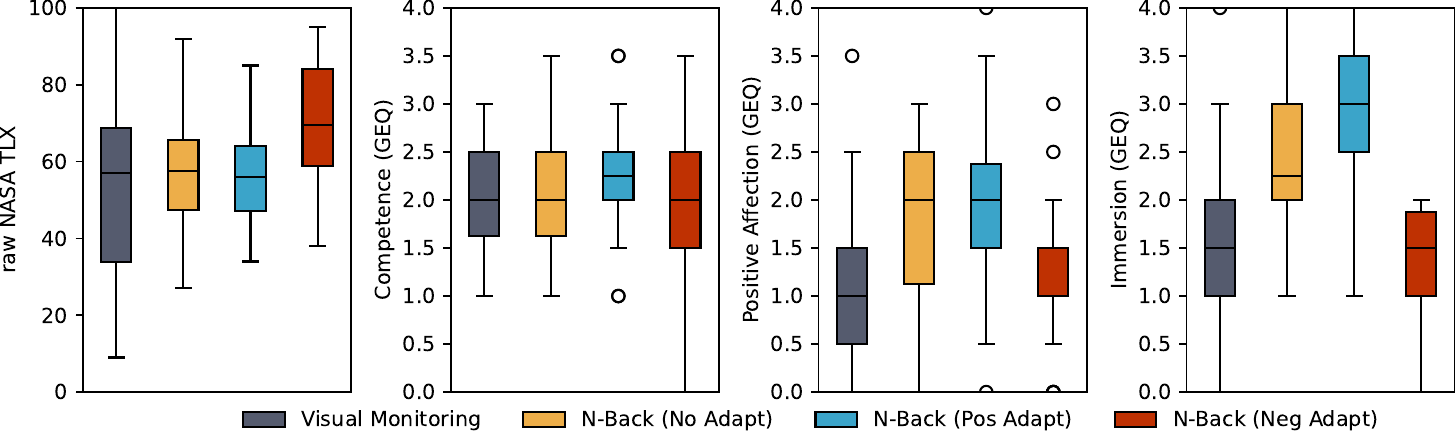}
    \caption{Subjective Results. Box-plots for perceived workload (NASA-TLX) and engagement (GEQ). Participants reported significantly more workload in the N-Back task with Negative Adaptation. Regarding perceived engagement, we found that participants experienced more Positive Affection and Immersion in N-Back (No Adapt) and N-Back (Pos Adapt) as compared to the Visual Monitoring task and the N-Back task in the Negative Adaptation.}
    \label{fig:results_nasa}
\end{figure}

\begin{figure*}[htbp]
    \centering
    \subfloat[Positive Stream variation for a representative participant]{%
        \includegraphics[width=.75\linewidth]{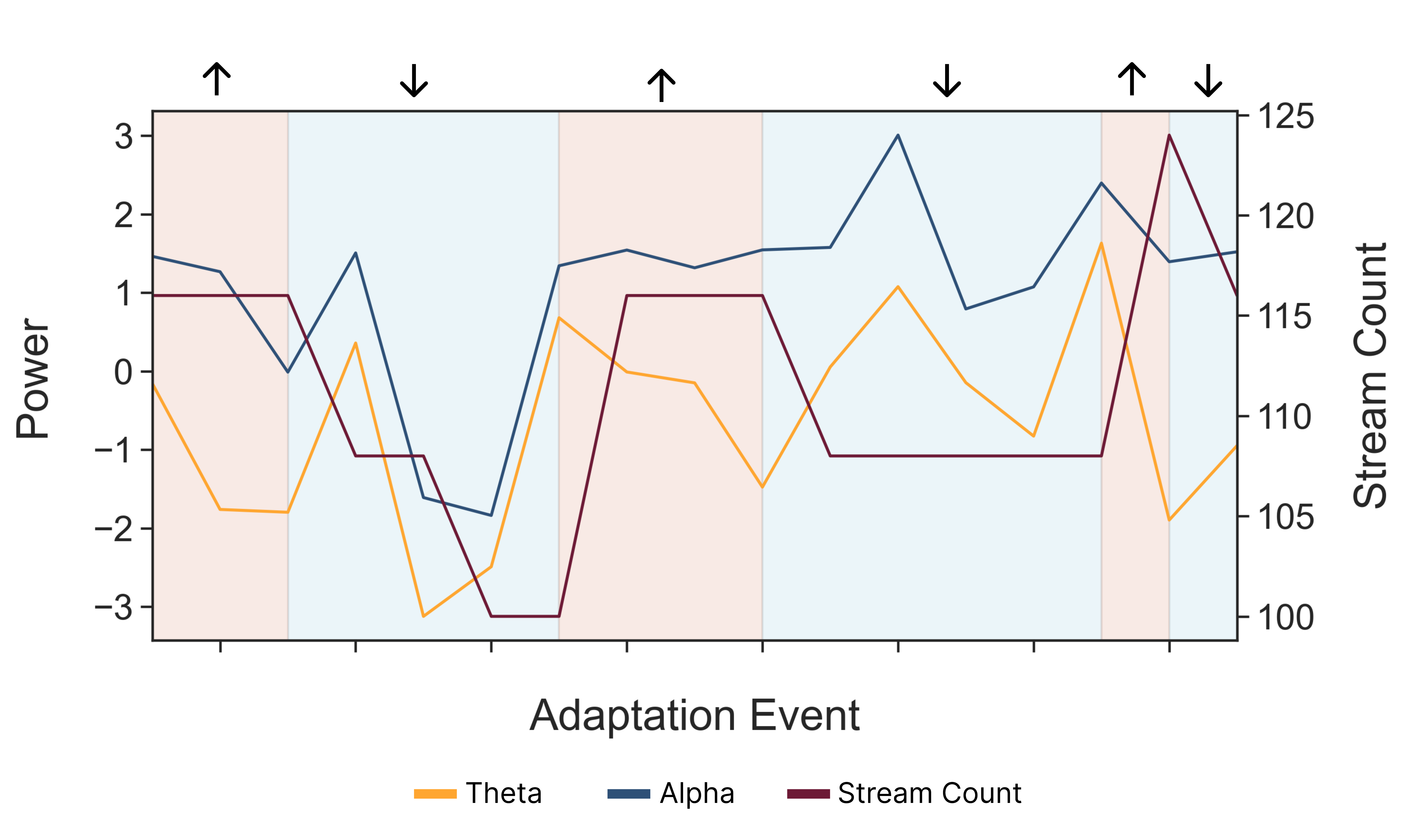}
        \label{fig:good_ada_stream}
    }
    
    \subfloat[Negative Adaptation Stream variation for a representative participant]{%
        \includegraphics[width=.75\linewidth]{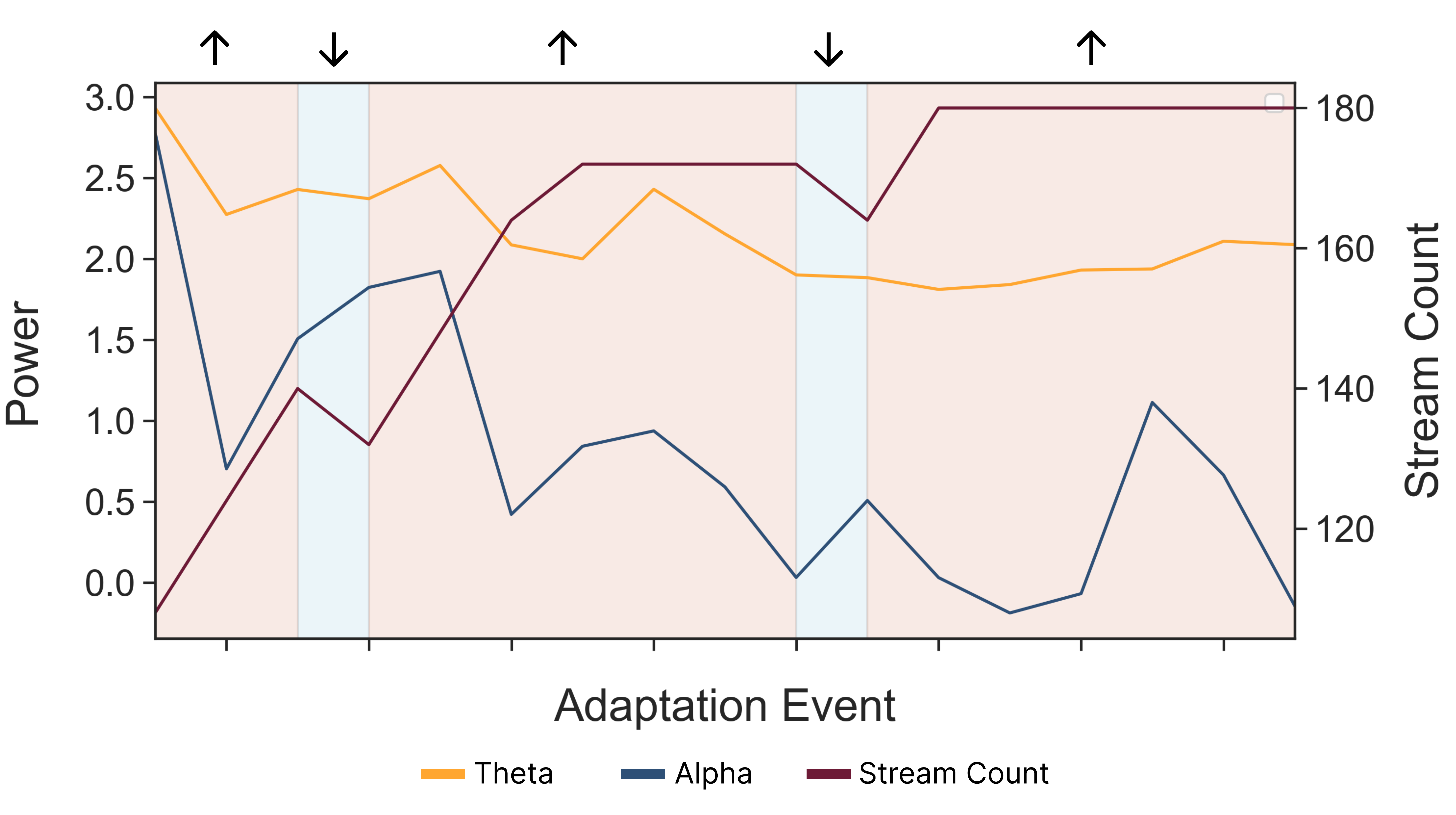}
        \label{fig:bad_ada_stream}
    }
    \caption{(a) Positive Stream variation and (b) Negative Adaptation Stream variation for representative participants. Yellow and Blue lines indicate the normalized Theta and Alpha frequency bands, while the dark red line represents the Stream Variation. Colored areas indicate whether the system increased (light red) or decreased (light blue) the NPCs Stream in a 20s time window. On top of each plot, the Stream increase is depicted by an arrow pointing up ( ${\uparrow}$ ), while if the Stream decreases, the arrow points down ( ${\downarrow }$ ).}
\end{figure*}

\subsubsection{Alpha}
The normality of Alpha power was assessed using the Shapiro-Wilk test, which indicated that the data were normally distributed ($W=0.98$, $p=.18$). A repeated measures ANOVA was conducted to examine the effect of \textsc{Block} on \textsc{Alpha}. A Repeated Measures ANOVA did not show any significant differences ($F=.45$, $p=.71$) as depicted on the left in \autoref{fig:results_eeg}.

\subsubsection{Theta}
As Theta power was not normally distributed (Shapiro-Wilk, $W=0.96$, $p=.02$),  we conducted a Friedman test indicating that the average Theta power did not change significantly across the different \textsc{Blocks} ($\chi=1.28$, $p=.73$) as depicted on the right in \autoref{fig:results_eeg}.

\subsection{Behavioral Results}

\subsubsection{Accuracy}
Shapiro-Wilk test showed a not normal distribution of accuracy scores ($W=0.96$, $p=.04$). We tested the effect of \textsc{Block} on Accuracy via a Friedman's test. We found a significant main effect ($\chi=27.36$, $p<. 001$). Post hoc comparisons with Bonferroni correction revealed that the mean accuracy in \textsc{Positive Adaptation}  ($\chi=1.28$, $p=.73$) ($M = .88$, $SD = .06$) was significantly increased from the mean score for \textsc{Negative Adaptation} ($M = .74$, $SD = .07$), $p < 0.01$. Additionally, the accuracy in \textsc{Negative Adaptation } was significantly lower as compared to the \textsc{N-Back} Block with no distractors ($M = .88$, $SD = .06$), $p < 0.01$. Results are depicted in \autoref{fig:results_acc}.

\subsubsection{Reaction Times}
We fitted a GLMM using REML and a nloptwrap optimizer on raw correct RTs with \textsc{Block} (N-Back No Adaptation, N-Back Positive Adaptation, N-Back Negative Adaptation) as a fixed effect and participant and amount of visual distractors per trial as a random effect.  We selected formula \texttt{rt $\sim$ Block +(1|participant) + (1|distractor)}. We removed outliers by excluding values exceeding three standard deviations above the mean~\citep{berger2021comparison}. However, we did not report any significant results. See \autoref{fig:results_rt}b.

\subsection{Subjective Results}

\subsubsection{Perceived Workload}

As Shapiro-Wilk showed a normal distribution ($W=0.99$, $p=.95$), an ANOVA showed that average raw NASA-TLX scores were significantly influenced by \textsc{Block} ($F=4.21$, $p<.001$). Pairwise comparisons via a Conover test with Bonferroni correction mimicked results on the accuracy, showing that \textsc{Negative Adaptation} resulted in a significantly higher workload ($M =70.13 $, $SD =16.97$) than \textsc{Positive Adaptation} ($M =57 $, $SD =13.09$), \textsc{N-Back} ($M =57.65 $, $SD =16.94$) and \textsc{Visual Monitoring} ($M =54.81 $, $SD =27.21$), all  $p < .001$. No significant differences were detected in other comparisons. Results are shown in \autoref{fig:results_nasa}.

\subsubsection{GEQ-Competence}
The Shapiro-Wilk normality test indicated a not normal distribution for the GEQ competence scores ($W=.95$, $p=.001$). A Friedman's test revealed no significant effects ($\chi=.51$, $p=.91$), see \autoref{fig:results_nasa}.

\subsubsection{GEQ-Positive Affection}
As the Shapiro-Wilk test showed a not parametric distribution ($W=.95$, $p=.002$), a Friedman rank sum test was conducted to examine the effect of \textsc{Block} on the GEQ Positive Affection scores. The analysis revealed a significant main effect of \textsc{Block} on GEQ Positive Affection scores ($\chi^2=26.23$, $p<.001$). Pairwise comparisons with a Conover test using a Bonferroni correction showed that GEQ Positive Affection scores in \textsc{Negative Adaptation} were significantly lower ($M=1.11$, $SD=.79$) than  \textsc{Positive Adaptation} ($M=1.86$, $SD=.98$) and \textsc{N-Back} ($M=1.86$, $SD=.99$), all $p<.001$. Identical results were found in the comparisons with the \textsc{Visual Monitoring} task, in which participants reported significantly lower subjective positive affection ($M=1.13$, $SD=.97$) than in the \textsc{Positive Adaptation} and \textsc{N-Back}. No differences were detected in the comparison between \textsc{Positive Adaptation} and \textsc{N-Back}. Results are depicted in \autoref{fig:results_nasa}.

\subsubsection{GEQ-Immersion}

The Shapiro-Wilk test indicated that the GEQ Immersion scores distribution of GEQ Immersion scores was non-parametric ($W=.95$, $p=.002$). A Friedman rank sum test revealed a significant main effect of \textsc{Block} ($\chi^2=34.2$, $p<.001$). Pairwise comparisons showed that the GEQ Immersion scores in \textsc{Negative Adaptation} ($M=1.36$, $SD=.1.36$) were significantly lower  than those in \textsc{Positive Adaptation} ($M=2.98$, $SD=.83$) and in \textsc{N-Back} ($M=2.48$, $SD=.71$), all $p<.005$. Visual Monitoring task condition showed significantly lower Immersion scores ($M=1.52$, $SD=.96$) as compared to \textsc{Positive Adaptation} and \textsc{N-Back} ( $p<.005$). No differences were detected in the comparison between \textsc{Visual Monitoring} and \textsc{Negative Adaptation}, see \autoref{fig:results_nasa}.

\subsection{Classification}
We evaluated the performance of a Linear Discriminant Analysis (LDA) model for predicting internal and external attention. We used EEG data from the Visual Monitoring for the External Attention label and the N-Back No Adaption for the Internal Attention label. We divided the dataset into training and validation sets in an 80/20 ratio, using a participant-wise split. The reported results include the accuracies and F1 score obtained on the test set.

\subsubsection{Feature Extraction}
We extracted EEG features based on the Power Spectral Densities (PSD) via Welch's method. We computed averaged alpha and theta based on the individual frequency range computed (see \autoref{sec:IAF_comp}) and delta (0.5 - 4 Hz), beta (13 - 30 HZ) and gamma (30-45 Hz) based on the preprocessing pipeline described in \autoref{sec:EEG_pipeline}. All frequency values were normalized based on the Resting state data. We used electrodes chosen for adaptation for alpha and theta as in \autoref{Adaptation_system}. For beta, we used the same frontal electrodes as theta \citep{putman2014eeg}, while for delta and gamma, we based our choice on previous work in internal-external attention classification \citep{vortmann2019eeg, vortmann2021, harmony1996eeg, darvas2010}. The EEG features were computed on 20s intervals, mirroring the time window used for adaptation.

\subsubsection{Classification Accuracy}The LDA model was trained on data from a subset of participants (N=$12$) and validated on data from a separate set (N=$5$). We then evaluated the model on the remaining participants (N=$5$). The LDA model achieved a training accuracy of $.8$ and a validation accuracy of $.76$ when using alpha, theta, beta, delta, and gamma measures to predict internal or external attention. While we report an accuracy of $.86$ and an F1 score of $.86$ on the test data. To understand which features were most informative for predicting internal and external attention, we examined the weight coefficients of the LDA model. The coefficients indicate the relative influence of each feature in predicting attention. Specifically, a positive coefficient for a feature indicates that higher values of that measure are associated with predicting external attention. 
In contrast, a negative coefficient indicates that higher feature values are associated with predicting internal attention. Our results showed that the alpha measure was predictive of external attention, with a positive coefficient of $.372$.
Delta power was majorly predictive of internal attention, with a negative coefficient of $-1.04$. The coefficients for the theta, beta, and gamma measures were $-.681$, $.281$, and $-.054$, respectively. These results suggest that alpha was specifically informative for external attention prediction, while delta and theta were indicative for internal attention. 

\section{Discussion}
We presented a physiologically adaptive VR system that employed EEG correlates of internal and external attention to perform dynamic visual complexity adjustments to enhance task performance.
We evaluated the effect of visual complexity adaptations, in the form of NPCs, on task performance, Alpha and Theta power, subjective workload, and engagement. In the study, participants performed a VR N-Back task recruiting WM resources. Here, we discuss our results regarding the outcome of our adaptive algorithms for modelling internal and external attention. Then, we envision applications for online attentional state detection and classification in VR and reflect on limitations and future work.  

\subsection{Internal and External Attention Modelling}
When users engage in VR tasks that feature both external and internal processing components, we hypothesized that they could benefit from an adaptation that could adjust the number of visual distractors in real-time to optimize their attentional state and enhance task performance. To achieve this, we designed two adaptive VR systems based on EEG alpha and theta power to optimize external and internal attentional states, respectively.

We identified four initial hypotheses. \textbf{HP1} and \textbf{HP2} predicted that the adaptive system designed for internal attention would improve WM task performance, while the system designed for external attention would decrease task performance. Our findings supported those two hypotheses, showing that participants performed better on the visual WM task when the adaptive system optimized distractors based on internal attention (\textbf{HP1}) and performance decreased when external attention was optimized (\textbf{HP2}). These results are consistent with previous research showing that attentional resources are essential for successful WM performance as the balance between external and internal attention can significantly affect task performance \citep{myers2017prioritizing}. When we need to recall and manipulate visual information and ultimately perform decisions, adapting task-irrelevant visual information can improve our task performance.
Conversely, it could be argued that distracting information could be removed from the environment to optimize internal attention for improving task performance. However, our results show that adaptation of visual distractors based on internal attention states enhanced perceived engagement through positive affection and immersion, supporting \textbf{HP3}. Participants reported higher levels of engagement when the adaptive system optimized distractors based on internal attention. On the other side, they reported significantly lower levels when interacting with the \textsc{Negative Adaptation}. Incorporating real-time adaptation based on internal attention states into VR systems could lead to more effective and enjoyable user experiences when high-level cognitive processing is involved. Additionally, our findings highlight the importance of considering internal and external attention in designing VR systems. Optimizing one at the expense of the other may adversely affect overall user experience and performance. In fact, an increase in engagement could have impacted the increase in task performance. Positive affection, for example, has been shown to enhance the focus of attention \citep{rowe2007positive}.  Finally, we verified \textbf{HP4}, as participants reported significantly higher levels of perceived workload when interacting with the \textsc{Negative Adaptation} as compared to the \textsc{Positive Adaptation} and to the N-Back with no distractors.  This finding aligns with previous research showing that increased external attentional demands can lead to a higher perceived workload \citep{rissman2009effect, kajimura2016we}. The perceived workload might have been associated with the continuous need to actively filter out task-irrelevant information, which can interfere with the processing of relevant information and increase cognitive load.
Our results suggest that an adaptive system that prioritizes internal attention can enhance executive performance in a VR environment. In contrast, external attention optimization can have a detrimental effect.

The results of the classification suggest that reliable decoding of internal and external attentional states in VR settings is possible, replicating similar results derived from AR settings \citep{vortmann2019eeg, vortmann2021}. Specifically, the main features contributing to the classification were alpha for internal states and theta and delta for external states. We can therefore state to have verified \textbf{H5}.

This finding is consistent with previous research on the role of alpha as a regulatory frequency in the balance between internal and external attention. Previous work showed how alpha decreases in response to attention-demanding tasks \citep{klimesch2012alpha}. Similarly, the role of theta power for internal attention is in line with previous literature, reflecting the maintenance of internal cognitive processes and inhibition of distracting information \citep{sauseng2005shift}. More interesting is the relevance derived from delta frequency band. Delta has been interpreted to act as  a functional modulator of sensory afferences that can interfere with internal concentration \citep{harmony2013functional}. Moreover, delta is associated with dynamic switching between external and internal attention \citep{jiang2021changes}, supporting their role in the inhibition of ongoing processes that can interfere with task execution. Our results align with previous research on the role of these frequency bands in attentional processes, highlighting their importance for understanding the neural mechanisms underlying attention in immersive environments.

\subsection{Applications for Attention-Aware VR adaptive systems}
Our findings have implications for the design and implementation of VR adaptive systems that aim to optimize attentional resources during tasks that jointly require internal and external processing. 

\subsubsection{Optimizing Internal Attention}

We found that we can optimize internal attention and improve task performance compared to an adaptive system that can be optimized for external attention. Moreover, we report that the performance with distractors in the \textsc{Positive Adaptation} did not significantly differ from the task executed without distractors. This can be specifically relevant for three application fields: VR productivity and cognitive training in healthy and clinical populations.

Our work showed how internal attention optimization can support task performance when engaged in a WM task. In the context of a virtual office \citep{knierim2021nomadic}, users might be novices to the multitude of visual stimuli, representing the surroundings or human colleagues, i.e. VR human avatars and more prone to distractions and inefficient workflows. Additionally, NPCs function can be adapted to provide cues, prompts, and reminders that can aid users in maintaining their focus and concentration on the task at hand. The design of a system that can minimize distraction while supporting engagement can be valuable for enhancing productivity in virtual environments, particularly in tasks that require working memory.

Cognitive training is another application where optimizing internal attention in WM tasks can be valuable. WM is essential in many cognitive tasks, such as problem-solving, decision-making, and learning, and is impaired in various clinical populations, including individuals with attention-deficit/hyperactivity disorder (ADHD) \citep{karbach2014making}. Cognitive training interventions aim to support WM performance while generalizing to other cognitive functions and have shown promising results in healthy, ageing and clinical populations, including ADHD \citep{cortese2016neurofeedback}.

In VR, adapting the system to optimize internal attention during cognitive training tasks could enhance the effectiveness of such interventions. Users could more efficiently engage in cognitive training tasks by minimizing distractions and improving focus, leading to better outcomes. Additionally, NPCs can be adapted to provide feedback, coaching, and reinforcement, enhancing cognitive training outcomes \cite{schroeder2020using}.

\subsubsection{Optimizing External Attention}

Internal and external attentional mechanisms play a crucial role in determining the effectiveness of VR applications. Even though the central purpose of attentional mechanisms is to facilitate the processing of relevant information over irrelevant one, sometimes internally directed attention can be undesired depending on the VR application and user state scenario. Internal attention might also refer not only to the prioritization of memory-related information but also to mind wandering \citep{gruberger2011towards} and rumination \citep{chuen2012impact}.

Therefore, if the user is engaged in a scenario where the visual information is task-relevant, but the user’s attention is internally directed, the VR system can increase the perceptual salience to capture the users’ attention or pause the interaction until they re-enter the external attention state. Such a scenario can be found in VR content or motor learning and visual analytics \citep{keim2008visual}, where users are provided with highly detailed and animated content. Such an interaction paradigm could prevent interrupting task-relevant thoughts and ignoring external information. This type of application could be based on the Optimal theory of learning \citep{wulf2016optimizing}, which postulates that an external focus of attention can result in improved learning skills compared to an internal one. Therefore optimizing for external attention in VR could allow for designing better learning systems.

\subsection{Limitations and Future Work}
We acknowledge that our work is prone to certain limitations related to the task we designed, their classification and how to improve our designed VR adaptive system.

In our study, we use the VR N-Back task, which inherently features an external shift of attention given its VR nature. This is an inherent limitation of using VR to recruit internal attention, and it must be acknowledged when designing experiments with a prominent visual component. To further evaluate the reliability of this paradigm, we suggest increasing the memory-related demands, such as increasing the amount of information held to be held in WM, i.e., moving from a 2-Back to a 3-Back VR task. Another possibility would be the addition of other internal components, such as episodic memory. Regarding the visual monitoring task, it is worth noting that while we did not explicitly verify whether participants directed their attention towards the NPCs, the task design and instructions provided to participants were based on prior research aimed at recruiting external attention \citep{cona2020theta, vitali2019action, arrabito2015sustained}. However, we acknowledge the limitation of not implementing a manipulation check based on eye-tracking. In future work, we plan to address this limitation by incorporating eye-tracking measures to assess participants' attentional focus accurately.

On the other hand, comparing the Visual Monitoring task to a VR version of an established external attention task, such as the visual oddball task \citep{putze2016}, would allow for better generalization of our results. These limitations and challenges are common in VR research, mainly when designing tasks that have to be ecologically situated.

Improving the generalizability of our results would support the reliability of our classification. Although we have selected tasks that theoretically recruit internal and external attention resources, our classifier could only discriminate between two tasks. Future work will address the training phase on more diverse tasks to validate our results further. Nonetheless, the high accuracy achieved in the between-participant task classification is comparable to previous work in AR \citep{vortmann2019eeg, vortmann2021}  and suggests the potential for online implementation to evaluate its performance.
Specifically, LDA is a machine learning model that allows for low computation and is successful for online cognitive state detection \citep{lotte2018review}. A new adaptation mechanism could be based on this classification approach, to balance the impact of multiple features and thus increase robustness against trial-to-trial variability.

Finally, our study demonstrated that conventional methods, such as the Welch periodogram computed on a moving time window, can adequately detect temporal variations in non-stationary signals. However, more advanced signal processing techniques like wavelet analysis can further improve the detection of temporal changes \citep{hillebrand2016direction}. Thus, implementing and evaluating wavelet analysis in future research may enhance the accuracy of attentional state classification. It is worth noting that efficient wavelet computation algorithms are available, which can be used in real-time applications~\citep{khalid2009towards, xu2009fuzzy}.

\section{Conclusion}
In this work, we presented a VR adaptive system based on EEG correlates of internal and external attention to dynamically adjust visual complexity and support task performance in a WM task. Visual complexity adjustments based on alpha and theta bands allowed for modulation of task-irrelevant elements adaptation and increased WM task performance. Furthermore, we showed that successful classification of EEG data in a VR N-Back task based on internal and external attention is possible. Even with simple machine learning algorithms, the classifier could reliably predict offline the attentional state of the participant, allowing for future implementation in real-time adaptive systems. 

\section{Open Science}
We encourage readers to reproduce and extend our results and analysis methods. Our experimental setup, collected datasets, and analysis scripts are available on the Open Science Framework\footnote{\url{https://osf.io/ar4fs/?view_only=0f6d2658af7342639c996684f23ce536}}.

\section{Declaration of Competing Interest}
The authors declare that they have no known competing financial interests or personal relationships tat could have appeared to influence the work reported in this paper.

\section{CRediT authorship contribution statement}
\textbf{Francesco Chiossi} Conceptualization, Methodology, Investigation, Formal Analysis, Data Curation, Writing - original draft, Writing - review \& editing, Project Administration.

\textbf{Changkun Ou} Methodology, Data Curation, Visualization, Writing - review \& editing.

\textbf{Carolina Gerhardt} Methodology, Investigation.

\textbf{Felix Putze} Conceptualization, Methodology, Supervision, Writing - review \& editing.

\textbf{Sven Mayer} Conceptualization, Supervision, Project administration, Funding acquisition, Writing - review \& editing.




\bibliographystyle{elsarticle-harv} 
\bibliography{bibliography} 



\end{document}